\documentclass{ws-mplb}
\usepackage{amssymb,epsfig}

\begin{document}

\markboth{V. M. Villalba and R. Pino}{Energy spectrum of the ground state...}

%
\catchline{}{}{}{}{}
%

\title{ENERGY SPECTRUM OF THE GROUND STATE OF A TWO DIMENSIONAL 
RELATIVISTIC HYDROGEN ATOM IN THE PRESENCE OF A CONSTANT MAGNETIC FIELD}

\author{\footnotesize V\'ICTOR M. VILLALBA\footnote{Alexander von Humboldt
Fellow}}

\address{Institut f\"ur Theoretische Physik. Universit\"at Frankfurt \\
D-60054 Frankfurt am Main, Germany \\
Centro de F\'{\i}sica \\ \
Instituto Venezolano de Investigaciones Cient\'{\i}ficas, IVIC \\
Apdo 21827, Caracas 1020-A, Venezuela \\
villaba@th.physik.uni-frankfurt.de}

\author{RAMIRO PINO}

\address{Department of Chemistry, Rice University, \\
MS60, Houston, TX77005-1892, USA \\
rpino@ruf.rice.edu}

\maketitle

\begin{history}
\received{(received date)}
\revised{(revised date)}
\end{history}

\begin{abstract}
We compute the energy spectrum of the ground state of a 2D Dirac electron in 
the presence of a Coulomb potential and a constant magnetic field perpendicular 
to the plane where the the electron is confined. With the help of a mixed-basis 
variational method we compute the wave function and the energy level and show 
how it depends on the magnetic field strength.
We compare the results with those obtained numerically as well as in the 
non relativistic limit. 
\end{abstract}


\section{Introduction}

Over the past years the study of systems of non-relativistic electrons 
confined to a plane in an electromagnetic field
background has attracted much attention in view of possible applications. This
problem is of practical interest because of the technological advances in
nanofabrication technology that have made the creation of low dimensional
structures like quantum wells, quantum wires and quantum dots possible.\cite
{Chakraborty} The relativistic extension of this problem
has also turned out to be of importance in the description of 
quantum two-dimensional phenomena such as the quantum
Hall effect and high-temperature superconductivity.\cite{Prange} Different
condensed matter physics phenomena point to the existence of
(2+1)-dimensional with an energy spectrum determined by the Dirac equation
Hamiltonian.\cite{Schakel1,Schakel2} In particular, the degenerate planar
semiconductor with low-energy electron dynamics is assumed to admit an
adequate description in terms of the (2+1)-dimensional relativistic Dirac
theory.\cite{Schakel3} In conclusion, the study of physical effects occurring 
in (2+1) dimensional systems of charged particles in strong external fields is
an interesting problem from the theoretical point of view as well
from its practical applications\cite{Khalilov2,Khalilov,Chiang,Ho,Ho2}.

In order to analyze relativistic quantum effects in the presence of strong
electromagnetic fields one should be able to compute the Green function
or to find exact solutions of the Dirac equation. Regrettably, the Dirac 
equation is exact solvable only in a very restricted family
of electromagnetic configurations.\cite{Bagrov} 

The non relativistic two-dimensional Hamiltonian describing the Coulomb
interaction $-\frac{Z}{r}$, between a conduction electron and donor impurity
center when a constant magnetic $\vec{{B}}$ field is applied perpendicular
to the plane of motion has been thoroughly discussed in the 
literature.\cite{Villalba2,Villalba,villalba2,Taut1,Mustafa}
Despite its simple form their solutions can not be expressed in terms of
special functions. An analogous situation occurs when we deal with the
(2+1) Dirac equation, therefore one has to apply numerical and approximate
techniques in order to compute the energy spectrum and the corresponding
wave functions. 

In order to compute the energy spectrum of the relativistic hydrogen atom we have at 
our disposal different numerical and semi-analytic methods. The Kato and and 
Lehman-Maelhy methods have also  been applied to solve the Dirac equation without 
variational collapse\cite{Falsaperla}  A large finite-basis-set method has been applied 
in order to calculate in 3D relativistic and nonrelativistic binding energies of an electron in
a Coulomb field a magnetic field.\cite{Chen}

It is the purpose of the present article to compute the energy spectrum of
the ground state of the (2+1) Dirac equation in the presence of a Coulomb
interaction and a constant magnetic field perpendicular to the plane where
the electron moves. For this purpose we use a two-term mixed-basis variational
approach.\cite{villalba2} We also compute the non relativistic limit with the help of the
Pauli equation. Finally, we compare the results with those
obtained numerically with the help of the Schwartz method, \cite{Schwartz}
which is a generalization of the mesh point technique.

The article is arranged as follows: In Sec. 2 we write the (2+1) Dirac
equation in the presence a constant magnetic field and a Coulomb potential;
we construct a two-term mixed-basis in order to compute the energy spectrum.
In Sec. 3, we solve the Pauli equation and show how the energy spectrum
depends on the spin orientation. In Sec. 4, with the help of a mesh method
we solve the Dirac equation  numerically. We compare the numerical result with
those obtained in the non relativistic limit and with the the help of the
variational method. Finally we briefly discuss the validity of our approach.
and further applications.

\section{Dirac equation}

The covariant Dirac equation\cite{Khalilov,shishkin} in atomic units,
$\hbar=M=e=1$, \ in the presence of an external electromagnetic field $A_{\mu
}$ can be written as follows
\begin{equation}
\left(  \tilde{\gamma}^{\mu}(\frac{\partial}{\partial x^{\mu}}-\Gamma_{\mu
}-i\frac{A_{\mu}}{c})+c\right)  \Psi=0, \label{Diracc}%
\end{equation}
where the Dirac matrices $\tilde{\gamma}^{\mu}$ satisfy the commutation
relation $\left\{  \tilde{\gamma}^{\mu},\tilde{\gamma}^{v}\right\}
_{+}=2g^{\mu\nu}.$The metric tensor $g_{\alpha\beta}$ written in polar
coordinates $(t,\rho,\vartheta)$ has the form:
\begin{equation}
g_{\alpha\beta}=diag(-1,1,\rho^{2}), \label{4}%
\end{equation}
and $\Gamma_{\mu}$ are the spinor connections \cite{Brill}. The vector
potential $A^{\alpha}$ associated with a 2D Coulomb potential and a constant
magnetic field interaction is:

\begin{equation}
A^{\alpha}=(-\frac{Z}{\rho},0,-\frac{{B}\rho^{2}}{2}).\label{5}%
\end{equation}
Choosing to work in the diagonal tetrad gauge we have that the
$\tilde{\gamma}$ matrices take the form
\begin{equation}
\tilde{\gamma}^{\rho}=\gamma^{1},\ \tilde{\gamma}^{\vartheta}=\frac
{\tilde{\gamma}^{2}}{\rho},\ \tilde{\gamma}^{t}=\gamma^{0},%
\end{equation}
where $\gamma^{\mu}$ are the standard constant Dirac matrices, satisfying the
anticommutation relations $\left\{  \gamma^{\mu},\gamma^{\nu}\right\}
=2\eta^{\mu\nu}.$

In the diagonal tetrad gauge,\cite{shishkin} the spinor connections take the
form
\begin{equation}
\label{conex}
\Gamma_{0}=0,\ \Gamma_{1}=0,\ \Gamma_{2}=\frac{1}{2}\gamma^{1}\gamma^{2}.
\end{equation}
Since we are working in 2+1 dimensions, it is convenient to introduce the
following representation of the Dirac matrices in terms of the Pauli matrices%
\begin{equation}
\gamma^{1}=\sigma_{1},\ \gamma^{2}=\sigma_{2},\ \gamma^{0}=\sigma_{3}.
\label{rep}%
\end{equation}
The Dirac equation (\ref{Diracc}) expressed in the diagonal tetrad gauge
commutes with the operators $i\frac{\partial}{\partial t}$ and $-i\frac
{\partial}{\partial\vartheta}$, therefore the spinor $\Psi$ can be written as
\begin{equation}
\Psi(t,\mathbf{\rho})=\frac{1}{\sqrt{2\pi\rho}}\exp(-iEt+ik_{\vartheta
}\vartheta)\psi(\rho), \label{Psi}%
\end{equation}
with 
\begin{equation}
\psi=\left(
\begin{array}
[c]{c}%
\psi_{1}(\rho)\\
\psi_{2}(\rho)
\end{array}
\right),
\end{equation}
where the $\sqrt{\rho}$ factor has been introduced in order to eliminate the
spinor contribution $\gamma^{\mu}\Gamma_{\mu}$ in the Dirac equation
(\ref{Diracc}).
Substituting Eq.(\ref{Psi}) into Eq.(\ref{Diracc}) and taking into account 
Eq. (\ref{conex}), we obtain
\begin{equation}
\left[  \frac{E}{c}+\frac{Z}{c\rho}+i\sigma^{2}\partial_{\rho}+\sigma
^{1}(\frac{k_{\vartheta}}{\rho}+\frac{B\rho}{2c})+\sigma^{3}c\right]  \psi=0.
\label{three}%
\end{equation}
After substituting the Pauli matrices into Eq. (\ref{three}) we reduce our problem to the
following system of coupled differential equations
\begin{equation}
\frac{d\psi_{2}}{d\rho}+\left(  \frac{k_{\vartheta}}{\rho}+\frac{B\rho}%
{2c}\right)  \psi_{2}+(\frac{E}{c}+c+\frac{Z}{c\rho})\psi_{1}=0, \label{p1}%
\end{equation}%
\begin{equation}
\frac{d\psi_{1}}{d\rho}-\left(  \frac{k_{\vartheta}}{\rho}+\frac{B\rho}%
{2c}\right)  \psi_{1}-(\frac{E}{c}-c+\frac{Z}{c\rho})\psi_{2}=0. \label{p2}%
\end{equation}

In order to analyze the nature of the eigenvalues $k_{\vartheta}$ of the
$-i\frac{\partial}{\partial\vartheta}$ operator, we notice that the spinor
$\Psi$ expressed in the (rotating) diagonal tetrad gauge,  is related to the
Cartesian (fixed)  spinor  $\Psi_{c}$ by means of the 
transformation\cite{shishkin}:
\begin{equation}
\Psi=S(\vartheta)^{-1}\Psi_{c},\label{ese}%
\end{equation}
where the matrix \ $S(\vartheta)$ can be written as%
\begin{equation}
S(\vartheta)=\exp(-\frac{\vartheta}{2}\gamma^{1}\gamma^{2})=\exp
(-i\frac{\vartheta}{2}\sigma_{3}).
\end{equation}
Noticing that $S(\vartheta)$ satisfies the relation
\begin{equation}
S(\vartheta+2\pi)=-S(\vartheta)
\end{equation}
we obtain
\begin{equation}
\Psi(\vartheta+2\pi)=-\Psi(\vartheta)\label{psi},%
\end{equation}
and we have $k_{\vartheta}=N+1/2$, where $N$ is an integer number. The rotating
Dirac spinor $\Psi$ can be written in terms of \ $\Psi_{c}$ as:
\begin{equation}
\Psi=\left(
\begin{array}
[c]{c}%
e^{i\vartheta/2}\Psi_{1c}\\
e^{-i\vartheta/2}\Psi_{2c}%
\end{array}
\right)  ,\quad\mathrm{with\quad}\Psi_{c}=\left(
\begin{array}
[c]{c}%
\Psi_{1c}\\
\Psi_{2c}%
\end{array}
\right)  .
\end{equation}
Taking into account the relation (\ref{ese}) between $\Psi$ and $\Psi_{c}$ as
well as (\ref{Psi}) we readily obtain
\begin{equation}
\Psi_{c}(\mathbf{r})=\frac{1}{\sqrt{2\pi\rho}}\left(
\begin{array}
[c]{c}%
e^{i(k_{\vartheta}-1/2)\vartheta}\psi_{1}(\rho)\\
e^{i(k_{\vartheta}+1/2)\vartheta}\psi_{2}(\rho)
\end{array}.
\right)
\end{equation}

The system of equations (\ref{p1})-(\ref{p2}) does not admit exact solution in
terms of special functions,  therefore we proceed to compute energy
eigenvalues with the help of variational techniques.

In non relativistic quantum mechanics, variational methods provide a powerful
technique for the construction of approximate eigenvalues and eigenfunctions
and for calculations involving sums over the complete energy spectrum. The
variational method cannot be straightforwardly extended to the Dirac case
because the Hamiltonian associated with Eq. (\ref{Diracc}) is not bounded from
below. Any positive-energy eigenvalue can collapse into a negative-energy
eigenvalue as the basis set is increased or as the nonlinear parameters of the
basis are varied.\cite{Quiney}

The earliest attempts at the solution of the relativistic basis set problem
revealed a tendency for any variational approach to a bound-state of positive
energy to collapse into the negative-energy region of the spectrum. This
effect, which was first noticed by Kim,\cite{Kim} has since become known as
variational collapse or finite basis set disease.

Among the possible ways to circumvent
the problems related to a finite basis and the variational collapse, we have
chosen to work with a quadratic Dirac equation:

\begin{equation}
\left[  \frac{d^{2}}{d\rho^{2}}+\left(  \frac{E}{c}+\frac{Z}{c\rho}\right)
^{2}+i\sigma^{2}\frac{Z}{c\rho^{2}}-\left(  \frac{k_{\vartheta}}{\rho}%
+\frac{B\rho}{2c}\right)  ^{2}-c^{2}+\frac{k_{\vartheta}\sigma^{3}}{\rho^{2}%
}-\frac{B\sigma^{3}}{2c}\right]  \psi=0 \label{quadratic}%
\end{equation}
which can be obtained after squaring Eq. (\ref{three}). It is worth noticing
that Eq. (\ref{quadratic}) is a second  order differential equation and
therefore it does not present the problems associated with the variational collapse.

In order to solve Eq. (\ref{quadratic}) we propose a mixed-basis variational
approach that has been successfully applied in the non relativistic case as
well as in computing the energy eigenvalues of the (2+1) Klein-Gordon
equation. We introduce the following trial function
\begin{equation}
\psi=c_{1}\psi_{1}+c_{2}\psi_{2}=c_{H}\psi_{H}+c_{O}\psi_{O},\label{mixed}%
\end{equation}
where $\psi_{1}=\psi_{H}$ and $\psi_{2}=\psi_{O}$ are the corresponding
hydrogen and oscillator wave functions associated with the ground state,
$c_{O}$ and $c_{H}$ are constants to be calculated. It is worth mentioning
that our basis is not orthogonal under the inner product $\left\langle
\psi_{i}\mid\psi_{j}\right\rangle =\int_{0}^{\infty}\psi_{i}^{\dagger}\psi
_{j}d\rho.$ Substituting (\ref{mixed}) into (\ref{quadratic}), and performing
variation on the basis coefficients $c_{j}$, we readily obtain the following
matrix equation:
\begin{eqnarray}
  \left(  \left\langle \psi_{i}(\rho),\frac{d^{2}\psi_{j}(\rho)}{d\rho^{2}%
}\right\rangle +(\frac{Z^{2}}{c^{2}}-k_{\vartheta}^{2})A_{ij}-\frac{B}%
{2c}F_{ij}+k_{\vartheta}G_{ij}+i\frac{Z}{c}H_{ij}\right)  c_{j}\label{matriz}%
\\
+\left(  (\frac{m{B}}{c}-c^{2}+\frac{E^{2}}{c^{2}})\delta_{ij}-\frac{1}%
{4}\frac{{B}^{2}}{c^{2}}D_{ij}+\frac{2EZ}{c^{2}}C_{ij}\right)  c_{j} \
=0\nonumber
\end{eqnarray}
with
\begin{eqnarray}
A_{ij}   =\left\langle \psi_{i}(r)\left|  \frac{1}{\rho^{2}}\right|  \psi
_{j}(r)\right\rangle ,\ G_{ij}=\left\langle \psi_{i}(r)\left|  \frac
{\sigma_{ij}^{3}}{\rho^{2}}\right|  \psi_{j}(r)\right\rangle, \label{mat}\\
C_{ij}   =\left\langle \psi_{i}(r)\left|  \frac{1}{\rho}\right|  \psi
_{j}(r)\right\rangle ,\ H_{ij}=\left\langle \psi_{i}(r)\left|  \frac
{\sigma_{ij}^{2}}{\rho^{2}}\right|  \psi_{j}(r)\right\rangle, \nonumber\\
D_{ij}   =\left\langle \psi_{i}(r)\left|  \rho^{2}\right|  \psi
_{j}(r)\right\rangle ,\ F_{ij}=\left\langle \psi_{i}(r)\left|  \sigma_{ij}%
^{3}\right|  \psi_{j}(r)\right\rangle. \nonumber
\end{eqnarray}
Instead of computing all the matrix terms in (\ref{mat}) one can use that
$\psi_{H}$ and $\psi_{O}$ are the corresponding hydrogen and oscillator wave
functions associated with the ground state
\begin{equation}
H|\Psi>=c_{H}E_{H}|\Psi_{H}>+c_{O}E_{O}|\Psi_{O}>.\label{ham}%
\end{equation}
Using the expression (\ref{ham}) we reduce Eq. (\ref{matriz}) to the following
matrix equation:
\begin{equation}
Q_{ij}c_{j}=0\label{matrix}%
\end{equation}
where the components of the square matrix $Q_{ij}$ are
\begin{equation}
Q_{11}=\left(  \frac{E}{c}\right)  ^{2}+\frac{2(E-E_{H})Z}{c^{2}}%
C_{11}-\left(  \frac{E_{H}}{c}\right)  ^{2}-D_{11}\left(  \frac{B}{2c}\right)
^{2}-\frac{B}{2c}(k_{\vartheta}+G_{11})%
\end{equation}%
\begin{eqnarray}
Q_{12}=\left[  \left(  \frac{E}{c}\right)  ^{2}-\left(  \frac{E_{H}}%
{c}\right)  ^{2}-\frac{Bk_{\vartheta}}{2c}\right]  S+\frac{2(E-E_{H})Z}{c^{2}%
}C_{12}-D_{12}\left(  \frac{B}{2c}\right)  ^{2}%
\end{eqnarray}%
\begin{equation}
Q_{22}=\left(  \frac{E}{c}\right)  ^{2}-\left(  \frac{E_{O}}{c}\right)
^{2}+\frac{2EZ}{c^{2}}C_{22}+\left(  \frac{Z}{c}\right)  ^{2}A_{22}%
\end{equation}
where $S=<\Psi_{H}|\Psi_{O}>$ is the overlap between the hydrogen and
oscillator wavefunctions.

The equation $\det(Q_{ij})=0$ gives as a result an algebraic equation for $E$ that
permits one to obtain the variational estimate for the energy. Among the
solutions of the secular equation we identify the energy as the lowest
positive root.

The advantage of the two-term mixed-basis variational approach is two-fold:
First,  it gives accurate results in the pure-Coulomb as well as in the strong 
magnetic field regimes and very good estimates for the intermediate region. 
Second, the variational wave function can be calculated in a closed form. We do
not have to handle large terms variational basis and the problems related
to balancing small and large components of the Hamiltonian. 

In this paper we are interested in computing the energy spectrum of the ground
$s$ state. The lowest energy configuration corresponds to a state that in the
absence of the Coulomb interaction has the electron spin directed opposite to
the magnetic field.\cite{Bagrov}

The solution of the Dirac equation in the presence of a constant magnetic
field with the electron spin directed opposite to $B$ is
\begin{equation}
\psi_{0}=\sqrt{\frac{B}{c}}\left(
\begin{array}
[c]{c}%
0\\
\sqrt{\rho}%
\end{array}
\right)  e^{-\frac{B\rho^{2}}{4c}},\label{psio}%
\end{equation}
the energy associated with spinor (\ref{psio}) does not depend on the magnetic
field strength and it is given by the simple expression:
\begin{equation}
\frac{E_{O}^{2}}{c^{2}}=c^{2},\label{energyo}%
\end{equation}
which corresponds to a zero non relativistic energy.

In order to compute the spinor $\psi_{H}$, solution of the (2+1) Dirac
equation in the presence of the Coulomb interaction we solve Eqs.
(\ref{p1})-(\ref{p2}) for $B=0$. The solution can be readily obtained in terms
of Laguerre Polynomials that \ for the $s$ state take the simple form
\begin{equation}
\psi_{H}=\frac{(2\lambda)^{{\nu}+\frac{1}{2}}}{\sqrt{\Gamma
(2{\nu}+1)}}\rho^{{\nu}}e^{-\lambda\rho}\left(
\begin{array}
[c]{c}%
\sqrt{1-E_{H}/c^{2}}\\
\sqrt{1+E_{H}/c^{2}}%
\end{array}
\right),  \label{psih}%
\end{equation}
where $\nu$ and $\gamma$ are given by
\begin{equation}
\lambda=\sqrt{c^{2}-\frac{E_{H}^{2}}{c^{2}}}\label{lambda}%
\end{equation}%
\begin{equation}
{\nu}=\sqrt{k_{\vartheta}^{2}-\frac{Z^{2}}{c^{2}}}.\label{gamma}%
\end{equation}
The energy of the ground state is
\begin{equation}
\frac{E_{H}}{c}=c\left[  \sqrt{1+\left(  \frac{Z}{c{\nu}}\right)
^{2}}\right]  ^{-1}.%
\end{equation}
Some coefficients of the matrix $Q_{ij}$ are divergent in the hydrogen-oscillator
basis. In order to circumvent this problem we
replace $\rho^{1/2}$ by $\rho^{1-{\nu}}$ in Eq. (\ref{psio}). This substitution
permits one to compute the coefficients $A_{ij}$ and does not sensitively affect 
the computation of the energy energy spectrum.

\section{Non relativistic limit}

In this section we are interested in computing the energy energy spectrum of
the 2+1 Coulomb atom in the presence of a constant magnetic field when one
consider spin effects. The non relativistic limit of the Dirac equation (\ref
{Diracc}) is given by the Pauli equation:
\begin{equation}
H\Psi =\left[ \frac{1}{2}(\vec{P}+\frac{1}{2c}\vec{B}\times \vec{r})^{2}-%
\frac{Z}{\rho }+\vec{s}\cdot \frac{\vec{B}}{c}\right] \Psi,  \label{Pauli}
\end{equation}
where ${\vec{s}}$ is the spin operator. For a constant magnetic field, we
have that ${\vec{s}}\cdot \vec{B}$ can be written as
\begin{equation}
\vec{s}\cdot \vec{B}=\frac{1}{2}\sigma _{3}B.
\end{equation}
Since we are dealing with a two-dimensional problem, we choose to work in
polar coordinates $(\rho ,\vartheta )$. The angular operator $-i\partial
_{\vartheta }$ commutes with the Hamiltonian (\ref{Pauli}), consequently we
can introduce the following {\it ansatz} for the eigenfunction:
\begin{equation}
\Psi =\frac{e^{im\varphi }}{\sqrt{2\pi }}\frac{u(\rho )}{\sqrt{\rho }}.
\label{fun}
\end{equation}
Substituting (\ref{fun}) into (\ref{Pauli}), we readily obtain that the
radial function $u(\rho )$ satisfy the second order differential equation
\begin{equation}
-\frac{1}{2}\frac{d^{2}u}{d\rho ^{2}}+\left[ \frac{1}{2}(m^{2}-\frac{1}{4})%
\frac{1}{\rho ^{2}}+\frac{B^{2}\rho ^{2}}{8c^{2}}-\frac{Z}{\rho }\right] u=%
\left[ E-\frac{B}{c}s-m\frac{B}{2c}\right] u.  \label{Paulir}
\end{equation}
The solution of equation (\ref{Paulir}) cannot be expressed in terms of
special functions.\cite{Bagrov} Taking into account how the spin couples to
the magnetic field in the Pauli equation, we can compute the energy
eigenvalues of Eq. (\ref{Paulir}). The energy spectrum of the Pauli can be
computed in a straightforward manner with the help of the solutions of the
Schr\"{o}dinger equation. The presence of the spin introduces a shift in
the energy proportional to the magnetic field strength.
\begin{equation}
E_{nonrel}=E_{Schr}\pm \frac{B}{2c},  \label{energyp}
\end{equation}
where the upper and lower signs correspond to the a magnetic field oriented
along or in opposite direction to the electron spin respectively. Since we
are interested in a configuration where the spin is opposite direction to the
spin, which corresponds to the lowest energy configuration, we choose the
negative sign in Eq. (\ref{energyp}).

The computation of energy spectrum of the (2+1) Schr\"{o}dinger equation in
the presence of the a Coulomb potential and a constant magnetic is a problem
that has been widely discussed in the literature and different analytic and
numerical techniques are at our disposal.\cite{Mustafa}

\section{Discussion of the results}

In this section we proceed to compute  the energy spectrum of the Dirac
equation with the help of the methods discussed in Sec. II and III. In order
to compare the accuracy of the variational approach as well as of the
non relativistic limit, the numerical computations of the relativistic energy
spectra are carried out with the help of the Schwartz method\cite{Schwartz}%
, which is a generalization of the mesh point technique for numerical
approximation of functions.

This method gives
highly accurate results given a thoughtful choice of the reference
function. For Eqs.(\ref{p1})-(\ref{p2}) we chose as the interpolation function
\begin{equation}
f(\rho) = \sum_m f_m \frac{u(\rho)}{(\rho-r_m)a_m},
\end{equation}
where
\begin{equation}
 u(\rho)= \sin[ \pi(\rho/h)^{1/2} ]
\end{equation}
$r_m$ is a zero of $u(\rho)$, $a_m$ is a zero of its derivative,
and $h$ is the step of the quadratic mesh.
The use of this scheme on Eq. (\ref{p1})-(\ref{p2}) leads
to an algebraic eigenvalue problem, giving a non-symmetric matrix
to be diagonalized in order to obtain the energy values. This selection gives
as result an overall convergence rate which goes approximately as $\epsilon
\approx 10^{(-2/3)N)}$ where N is the number of truncations.
The accuracy
of the technique has been verified by computing the energy spectrum of the 2D
Hydrogen atom\cite{Villalba2,Villalba,villalba2} and reproducing  the analytic results
obtained by Taut\cite{Taut1} for the excited states.

\begin{figure}
\epsfig{file=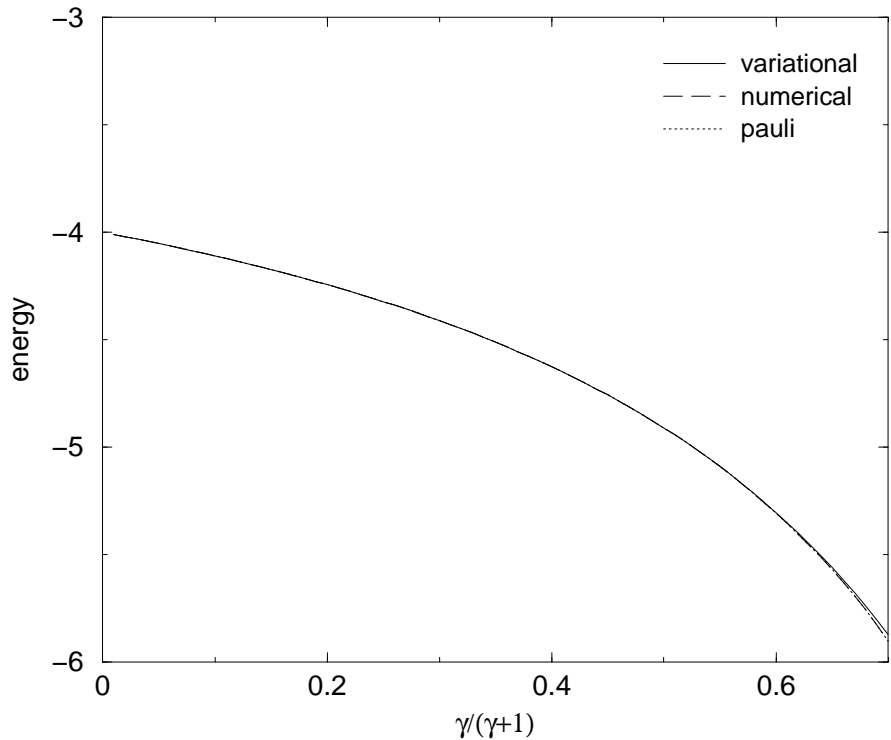,height=10cm}
\caption{Energy $E-c^2$  in atomic Rydberg units of the $1s^-$ state as a function of
$\protect\gamma^{\prime
}=\gamma/(\gamma+1) $. The long-dashed line is obtained by numerical methods;
the solid line corresponds to the mixed-basis variational method.
The dotted line is obtained by using the Pauli equation}.
\end{figure}

Fig. 1 shows the dependence of the non relativistic energy $E-c^2$ on the
magnetic field strength $B$. In order the compare the results with those
previously reported in the literature,\cite{Villalba2,Mustafa} we introduce the
parameter $\gamma=B/c$. Fig. 1 shows that the two-term variational method gives
very good results
for $\gamma ^{\prime }=\gamma/(\gamma+1)<0.7$ The non relativistic approximation works for all values of
the magnetic field strength. Table \ref{tab:table1} shows clearly that the 
variational approach is better
suited for small and intermediate values of $\gamma ^{\prime }$ but the
non relativistic Pauli equation guarantees at least four figures for
any value of $B$ and gives reliable results in
a range of values that its not restricted to values $B<<B_{cr}$ as the
semiclassical approximation.\cite{Ho}

For weak magnetic fields, Ho and Khalilov\cite{Ho} have obtained
approximate energy values with the help of a semiclassical method, their
result, written in atomic Rydberg units,  takes the form:
\begin{equation}
E_{sem}=2\left( c^{2}+\frac{\gamma k_{\vartheta }}{2}\right) \left[ 1+\frac{1%
}{c^{2}(n+\sqrt{k_{\vartheta }^{2}-1/c^{2}})^{2}}\right] ^{-1/2}
\label{khal}
\end{equation}
Expression (\ref{khal}) gives good energy estimates for $\gamma ^{\prime
}<0.05$. In fact, for $\gamma ^{\prime }=0.05$ one obtains $E_{sem}=-4.02649$%
, a value slightly higher than that one obtained numerically and with the help of
variational techniques. For strong magnetic fields the authors obtain
recursion relations that determine the coefficients of the series expansion
of the wave function, the possible energies, and the magnetic fields,
showing that the two dimensional Dirac electron in the presence of an
homogeneous 
magnetic field is a quasiexactly solvable problem. Unfortunately, just like in
the non-relativistic\cite{Taut1} and Klein-Gordon\cite{Villalba} cases, the 
quasiexactly solvable energies do not correspond to the ground state. 

Since the energy spectrum for
the ground state,
in the pure magnetic  field regime does not depend on $B$ (\ref{energyo}), 
the variational 
energy slightly deviates from the numerical value for strong magnetic fields.
Regarding the methods and results presented in this article we have to
emphasize that we do not require any special identification between small
and large components of the Dirac spinor, a condition that is present in
most of the alternative variational methods available in the literature.
The electromagnetic configuration given by Eq.(\ref{5}) presents the interesting
feature of combining a Coulomb potential with a magnetic field in a way
that makes unsuitable a single trial-function approach. The accuracy of
the mixed-basis approach can be improved when one uses a many-terms basis,
nevertheless this one introduces higher order $Q_{ij}$ matrices and
further complications in the computation of the energy eigenvalues.

\begin{table}[pt]
\tbl{\label{tab:table1} Relativistic energy values $E-c^2$ in atomic Rydberg units 
for $s$ state and a
comparison with the non relativistic energy spectrum. The first column
corresponds to the value obtained via the mixed-basis variational method,
the second column is the numerical value
relativistic energy $E-c^{2}$, the third column corresponds to the energy
obtained after solving the Schr\"odinger, the fourth column is the energy
spectrum of the Pauli equation, the fifth column  (diff1) is the difference
between the the first and second columns, the sixth column (diff2) is difference
between the second a fourth column.}
{\scriptsize
\begin{tabular}{ccccccc} \toprule
$\gamma ^{\prime }$ & variational & numerical & Schr\"{o}dinger & Pauli
& diff1 & diff2 \\ \colrule
0.01 & -4.010308476 & -4.010303938 & -3.9999905 & -4.01009151 & -4.5385 $10^{-6}$ 
& 0.000212427 \\ 
0.05 & -4.052584016 & -4.052582214 & -3.9997404 & -4.052371979 & -1.80165 $10^{-6}$ 
& 0.000210235 \\ 
0.10 & -4.110158292 & -4.110161852 & -3.9988434 & -4.109954511 & 3.55978 $10^{-6}$ 
& 0.000207341 \\  
0.15 & -4.17375409 & -4.173760085 & -3.9970851 & -4.173555688 & 5.99508 $10^{-6}$ 
& 0.000204397 \\ 
0.20 & -4.244337352 & -4.244360501 & -3.9941592 & -4.2441592 & 2.31488 $10^{-5}$ 
& 0.000201301 \\ 
0.25 & -4.32313341 & -4.323172984 & -3.9896419 & -4.322975233 & 3.95743 $10^{-5}$ 
& 0.000197751 \\ 
0.30 & -4.411646466 & -4.411703859 & -3.9829381 & -4.411509529 & 5.73929 $10^{-5}$ 
& 0.00019433 \\ 
0.35 & -4.511776714 & -4.511854721 & -3.9732025 & -4.511664038 & 7.8007 $10^{-5}$ 
& 0.000190683 \\ 
0.40 & -4.625971588 & -4.626065158 & -3.9592116 & -4.625878267 & 9.35697 $10^{-5}$ 
& 0.000186891 \\ 
0.45 & -4.757387992 & -4.75752472 & -3.9391601 & -4.757341918 & 0.000136728
& 0.000182802 \\ 
0.50 & -4.91019842 & -4.910498134 & -3.9103193 & -4.9103193 & 0.000299714 &
0.000178834 \\ 
0.55 & -5.089955586 & -5.090843552 & -3.8684466 & -5.090668822 & 0.000887966 &
0.00017473 \\ 
0.60 & -5.303980092 & -5.306876432 & -3.806706 & -5.306706 & 0.00289634 & 
0.000170432 \\ 
0.65 & -5.561460262 & -5.570890138 & -3.7135808 & -5.570723657 & 0.009429876 & 
0.000166481 \\ 
0.70 & -5.871135116 & -5.902019599 & -3.5685234 & -5.901856733 & 0.030884483 & 
0.000162866 \\  \botrule
\end{tabular}}

\end{table}

Regarding the range of validity of variational and non relativistic approximation,
we have that both methods give very good results for values of $\gamma < 1$, which
correspond, for the hydrogen atom, to 
$B < 2.35 \times 10^9$ G, a value that is
far from the values of $B$ that one create in a laboratory\cite{Lai}.

However strong fields can be mimicked in nanomaterials like GaAs or InSb with
very small effective masses, where the methods and techniques of quantum effects
in the presence of strong fields can be applied \cite{Lai}. In  this physical scenario
the techniques developed in this article could be of help.

\section*{Acknowledgments}
One of the authors V.M.V acknowledges a Fellowship from
the Alexander von Humboldt Stiftung.

\end{document}